\documentclass[final]{siamltex}
\usepackage{graphicx}

\title{Deflated GMRES for Systems with Multiple Shifts and Multiple Right-Hand Sides\footnotemark[1]}

\author{Dean Darnell\footnotemark[2]
\and Ronald B. Morgan\footnotemark[3]
\and Walter Wilcox\footnotemark[4] }

\begin{document}
\bibliographystyle{plain}

\maketitle

\renewcommand{\thefootnote}{\fnsymbol{footnote}}
\footnotetext[1]{This work was partially supported by the National Science Foundation, Computational Mathematics Program under grant 0310573 and the National Computational Science Alliance.  It utilized the SGI Origin 2000 and IBM p690 systems at the University of Illinois.  The second author was also supported by the Baylor University Sabbatical Program.}
\footnotetext[2]{Department of Physics, Baylor
University, Waco, TX 76798-7316.  ({\tt Dean\_Darnell@baylor.edu}).}
\footnotetext[3]{Department of Mathematics, Baylor
University, Waco, TX 76798-7328 ({\tt Ronald\_Morgan@baylor.edu}).}
\footnotetext[4]{Department of Physics, Baylor
University, Waco, TX 76798-7316.  ({\tt Walter\_Wilcox@baylor.edu}).}
\renewcommand{\thefootnote}{\arabic{footnote}}

\begin{abstract}
We consider solution of multiply shifted systems of nonsymmetric linear equations, possibly also with multiple right-hand sides.  First, for a single right-hand side, the matrix is shifted by several multiples of the identity.  Such problems arise in a number of applications, including lattice quantum chromodynamics where the matrices are complex and non-Hermitian.  Some Krylov iterative methods such as GMRES and BiCGStab have been used to solve multiply shifted systems for about the cost of solving just one system.  
Restarted GMRES can be improved by deflating eigenvalues for matrices that have a few small eigenvalues.  We show that a particular deflated method, GMRES-DR, can be applied to multiply shifted systems.  

In quantum chromodynamics, it is common to have multiple right-hand sides with multiple shifts for each right-hand side.  We develop a method that efficiently solves the multiple right-hand sides by using a deflated version of GMRES and yet keeps costs for all of the multiply shifted systems close to those for one shift.  An example is given showing this can be extremely effective with a quantum chromodynamics matrix.
\end{abstract}



\pagestyle{myheadings}
\thispagestyle{plain}
\markboth{D. DARNELL, R. B. MORGAN and W. WILCOX}{DEFLATION FOR SHIFTED SYSTEMS}

\section{Introduction}

We consider the iterative solution of a large system of linear equations that not only has multiple right-hand sides, but also has multiple shifts for each right-hand side.  Let $nrhs$ be the number of right-hand sides and $ns$ be the number of shifts.  Then the problem is
\begin{equation}\label{eq:matrix}
(A - \sigma_i\*I)x_i^j = b^j,
\end{equation}
with $j=1,\ldots, nrhs$ and $i=1,\ldots,ns$.  Here $A$ is a large matrix which may be nonsymmetric or complex non-Hermitian.  We assume there is no preconditioning.  The shift $\sigma_1$ will be referred to as the base shift.  For the rest of the paper, we will leave off the superscripts when it is clear which right-hand side is being dealt with. 

Systems with multiple right-hand sides occur in many applications (see, for example,~\cite{FrMa} for some applications).  There are several applications that need solution of multiply shifted systems, for example, control theory~\cite{DaSa}, and time-dependent differential equations and frequency response computations~\cite{Fr93b}.  Some important problems in lattice quantum chromodynamics (lattice QCD) have both multiple right-hand sides and multiple shifts.  For example, the Wilson-Dirac formulation~\cite{Qcdconf2} and overlap fermion~\cite{NaNe,FrOverlap2} computations both lead to such problems.  Very large complex non-Hermitian matrices are needed.  For Wilson-Dirac matrices, the right-hand sides represent different noise vectors.  The shifts correspond to different quark masses that are used in an extrapolation.

A standard way to solve systems with multiple right-hand sides is to use a block approach~\cite{Sa96,bgdr}.  However, block methods can be costly due to the orthogonalization expense, and also storage requirements can be prohibitive.  Block methods are not generally used in lattice QCD.  Even if multiple right-hand sides are solved sequentially instead of together in a block, it is important to take advantage of the fact that several systems share the same matrix.  Information developed during the solution of one right-hand side can be used to help with others.  This idea has been developed in different ways, including in seed methods~\cite{ChWa} and with Richardson iteration~\cite{SiGa}.  Deflation methods can also be used (see~\cite{GMRES-DR, PadeStMaJoMa} and their references).  Deflation approaches appearing in QCD literature include~\cite{dF,EdHeNa,DoLeLiZh,NeEiLiNeSc}.  Deflation involves computing eigenvectors corresponding to the smallest eigenvalues (or other outstanding eigenvalues) and using them to remove the eigenvalues from the effective spectrum for the iterative method.  GMRES-DR~\cite{GMRES-DR} is a deflation method that has been adapted for multiple right-hand sides~\cite{gproj}.  Some details of GMRES-DR are given in the next section.

Krylov methods have been developed for shifted matrix problems.  Again, some details are in the next section.

The goal of this paper is to  develop deflated methods for dealing with multiple shifts along with both a single right-hand side and multiple right-hand sides.  Section 2 has background material on current methods for solving shifted systems and on deflated methods for multiple right-hand sides.  In Section 3, deflation is used for multiply shifted systems with a single right-hand side, or the first of several right-hand sides.  A multiply-shifted version of GMRES-DR is developed.  There is also some comparison of GMRES with FOM for multiple shifts.  Section 4 looks at solving the second and subsequent right-hand sides using eigenvector information and GMRES.  Then the next section deals with the case of related right-hand sides.

\section{Review}

\subsection{Krylov methods for shifted systems}

The Krylov subspace generated with the shifted matrix $A-\sigma_i I$ and with starting vector $r_0$ is 
\begin{equation}\label{eq:krylov}
Span\{r_0,(A-\sigma_i\*I)r_0,(A-\sigma_i\*I)^2 r_0,...,(A-\sigma_i\*I)^{m-1}r_0\}.
\end{equation}
Krylov subspaces are shift invariant in that the subspace is the same regardless of the choice of $\sigma_i$.  Therefore, one Krylov subspace can be used to solve several shifted systems as long as all the systems have the same right-hand side, or at least right-hand sides that are multiples of each other.  Note that we are assuming there is no preconditioning.  With a preconditioner, systems with different shifts would no longer have equivalent Krylov subspaces.
So for the non-preconditioned case, it is fairly straightforward to develop versions of nonrestarted Krylov methods for multiply shifted systems.  Such versions have been given for the conjugate gradient method~\cite{Fr90,YoVo}, nonrestarted GMRES~\cite{DaSa}, and both QMR and TFQMR~\cite{Fr93b}.  A multiply shifted version of BiCGStab has also been developed~\cite{Fr03}.  

Restarted Krylov methods are not as straightforward.  After a restart, all systems need to have parallel right-hand sides.  This means that all residual vectors formed at the end of a cycle of the Krylov method need to be multiples of each other.  For the FOM method~\cite{Sa96}, this happens automatically~\cite{Si03}.  For GMRES~\cite{SaSc,Sa96}, the residuals can be forced to be parallel, as shown by Frommer and Glassner~\cite{FrGl}.  
To see how this is possible, we need the Arnoldi recurrence~\cite{Sa96}:
\begin{equation}
 AV_m = V_{m+1} \bar H_m, \label{recurAr}
\end{equation} 
where $V_m$ is a $n$ by $k$ matrix whose columns span the Krylov subspace, $V_{m+1}$ is the same except for an extra column
and $\bar H_m$ is an upper-Hessenberg $m+1$ by $m$ matrix.  For a shifted system, this becomes 
\[
 (A-\sigma_i I)V_m = V_{m+1} (\bar H_m - \sigma_i \bar I),  
\] 
where $\bar I$ is the $m+1$ by $m$ identity matrix. 

Say that after a restart the approximate solution for the i$th$ shifted system is $\tilde x_{0,i}$ and the residual vector is $r_{0,i} = \beta_i r_{0,1}$.  So it is assumed that the residual is parallel to that of the base shift.  The system is then 
\[(A-\sigma_i I)(x_i- \tilde x_{0,i}) = r_{0,i} = \beta_i r_{0,1}.\] 
For the next cycle, let the base system have standard GMRES solution $V_m d_1$, where $d_1$ is the solution of $\min ||c -(\bar H_m - \sigma_i I)d_1||$, with $r_{0,1} = V_{m+1} c$.  Let $r_i$ be the new residual vector for the i$th$ shifted system after this cycle.  We need the approximate solution $V_m d_i$ to be chosen so that $r_i = \beta_i^{new} r_1$, for some scalar $\beta_i^{new}$.  So 
\begin{equation}
\beta_i r_{0,1} - (A-\sigma_i I) V_m d_i = \beta_i^{new} (r_{0,1} - (A-\sigma_1)V_m d_1). \label{gsh1}
\end{equation} 
After using the Arnoldi recurrence~(\ref{recurAr}), we have
\begin{equation}
V_{m+1}(\beta_i c - (\bar H_m - \sigma_i \bar I)d_i) = \beta_i^{new} V_{m+1}(c - (\bar H_m - \sigma_1 \bar I)d_1). \label{gsh2}
\end{equation} 
Next the $V_{m+1}$ can be dropped, and we let $s = c - (\bar H_m - \sigma_1 \bar I)d_1$ and rearrange:
\[(\bar H_m - \sigma_i \bar I) d_i = \beta_i c - \beta_i^{new} s.
\]
We now use a QR factorization, $\bar H_m - \sigma_i \bar I= QR$, with $Q$ being an $m+1$ by $m+1$ orthogonal matrix and $R$ being $m+1$ by $m$ upper-triangular, and get 
\[R d_i = \beta_i Q^T c - \beta_i^{new} Q^T s.
\]
The value of $\beta_i^{new}$ can be determined from the last row (note the left side of the equation has zero in the last row).  Then solution of an upper-triangular system determines $d_i$.

A shortcut formula for the residual norm of the $i$th shifted system is
\begin{equation}
||r_i|| = || \beta_i c - (\bar H_m -\sigma_i \bar I) d_i||.\label{gsh3}
\end{equation}
The new approximate solution is $\tilde x_i = \tilde x_{0,i} + V_m d_i$.  The systems then become 
$A(x_i- \tilde x_i) = \beta_i^{new} r_{1}.$  
For more details, see~\cite{FrGl}.  We will refer to this approach as GMRES-Shifts or GMRES-Sh.  Note that only the base system has the minimum residual property.  The solution of the other shifted systems is not equivalent to GMRES applied to those systems.  This is in contrast to FOM-Shifts~\cite{Si03}, which has each shifted system solved with the FOM approach.  Like FOM-Sh, GMRES-Sh uses the same polynomial (after shifting) for all shifted systems~\cite{FrGl,Si03}.

\subsection{Deflated GMRES}

Small subspaces for restarted GMRES can slow convergence for difficult problems.  Deflated versions of restarted
GMRES~\cite{GMRES-E,KhYe,ErBuPo,ChSa,Sa95B,BaCaGoRe,BuEr,LCMo,DS99,GMRES-IR,
GMRES-DR} can improve this, when the problem is difficult due to a few
small eigenvalues.  One of these approaches is related to Sorensen's
implicitly restarted Arnoldi method for eigenvalues~\cite{So} and is called
GMRES with implicit restarting~\cite{GMRES-IR}.  A mathematically
equivalent method, called GMRES with deflated restarting
(GMRES-DR)~\cite{GMRES-DR}, is also related to Wu and Simon's restarted
symmetric Lanczos eigenvalue method~\cite{WuSi}.  See~\cite{Arnoldi-R,St01,HRAM} for
some other related eigenvalue methods.  

We will concentrate on GMRES-DR~\cite{GMRES-DR}, because it is efficient and relatively
simple.  Approximate eigenvectors corresponding to the small eigenvalues are
computed at the end of each cycle and are put at the beginning of the next
subspace.  Letting $r_0$ be the initial residual for the linear equations
at the start of the new cycle and $\tilde y_1, \ldots
\tilde y_k$ be harmonic Ritz vectors~\cite{IE,Fr92,PaPavdV,IEN}, the
subspace of dimension $m$ used for the new cycle of GMRES-DR(m,k) is
\begin{equation}
Span\{\tilde y_1, \tilde y_2, \ldots \tilde y_k, r_0, A r_0, A^2 r_0, A^3
r_0, \ldots ,A^{m-k-1} r_0 \}. \label{ss}
\end{equation}
This can be viewed as a Krylov subspace generated with starting vector
$r_0$ augmented with approximate eigenvectors.  Remarkably, the whole
subspace turns out to be a Krylov subspace itself (though not with $r_0$ as
starting vector)~\cite{GMRES-IR}.
Once the approximate eigenvectors are moderately accurate, their inclusion
in the subspace for GMRES essentially deflates the corresponding
eigenvalues from the linear equations problem.  

GMRES-DR generates a recurrence similar to the Arnoldi recurrence~(\ref{recurAr}).  It is 
\begin{equation}
 AV_m = V_{m+1} \bar H_m, \label{recur0}
\end{equation} 
where $V_m$ is a $n$ by $m$ matrix whose columns span the subspace (\ref{ss}), $V_{m+1}$ is the same except for an extra column and $\bar H_m$ is an $m+1$ by $m$ matrix that is upper-Hessenberg except for a full $k+1$ by $k$ leading portion.  A part of this recurrence can be separated out to give
\begin{equation}
 AV_k = V_{k+1} \bar H_k, \label{recur1}
\end{equation} 
where $V_k$ is a $n$ by $k$ matrix whose columns span the subspace of
approximate eigenvectors, $V_{k+1}$ is the same except for an extra column
and $\bar H_k$ is a full $k+1$ by $k$ matrix.   This recurrence allows access to both the approximate eigenvectors and their products with $A$ while requiring
storage of only $k+1$ vectors of length $n$.  The approximate eigenvectors in GMRES-DR actually span a small Krylov subspace of dimension $k$. 

\subsection{Deflated GMRES for multiple right-hand sides}
 
The multiple right-hand side approach that will be adapted here for multiple shifts is called GMRES-Proj.  The first right-hand side is solved with GMRES-DR, then the eigenvector information thus generated is used to deflate eigenvalues from other right-hand sides~\cite{gproj}.  This is done with a simple minimum residual projection over the eigenvectors alternated with cycles of regular GMRES(m).  The expense of the projection does not generally add much to the GMRES cost.

The algorithm for the projection is given next.  It projects over the space of harmonic Ritz vectors spanned by the columns of $V_k$ in Equation~(\ref{recur1}). 
This requires only $3k+2$ vector operations of length $n$.  

\vspace{.10in}
\begin{center}
\textbf{Minres Projection for $V_k$}
\end{center}
\begin{enumerate}
 \item Let the current approximate solution be $\tilde x_0$ and the current system
of equations be $A(x-\tilde x_0) = r_0$.  Let $V_{k+1}$ and $\bar H_k$ be from Equation (\ref{recur1}).  
 \item Solve min$||c - \bar H_k d||$, where $c = (V_{k+1})^T r_0$.
 \item The new approximate solution is $\tilde x = \tilde x_0 + V_{k}d$.
 \item The new residual vector is $r = r_0 - AV_{k} d = r_0 - V_{k+1} \bar H_k d$.
\end{enumerate} 
\vspace{.15in}

The GMRES-Proj method that follows is for all right-hand sides except for the first one.  See~\cite{gproj} for more details.

\vspace{.10in}
\begin{center}
\textbf{GMRES(m)-Proj(k)}
\end{center}
\begin{enumerate}
 \item After applying the initial guess $\tilde x_0$, let the system
of equations be $A(x-\tilde x_0) = r_0$.  
 \item If it is known that the right-hand sides are related, project over the previous computed solution vectors.
 \item Apply the Minres Projection for $V_k$.  This uses the $V_{k+1}$ and $\bar H_k$ matrices developed while solving the first right-hand side with GMRES-DR.
 \item Apply one cycle of GMRES(m).
 \item Test the residual norm for convergence (can also test during the GMRES cycles).  If not satisfied, go back to step 3.
\end{enumerate} 
\vspace{.15in}

\section{Deflated GMRES with Multiple Shifts}

Subsection 2.1 gave some details of how restarted GMRES can be implemented in order to simultaneously solve multiply shifted systems.
For the deflated restarted GMRES method GMRES-DR, we can also solve multiply shifted systems concurrently.  The key is that GMRES-DR has subspaces that are Krylov subspaces (as mentioned in Section 2.2), and they always contain the current right-hand side vector.  The derivation in Subsection 2.1 for solving multiply shifted systems with GMRES also applies here with very slight change.  Because GMRES-DR has Krylov subspaces, it has the Arnold-like recurrence~(\ref{recur0}) which can be used in place of GMRES's Arnoldi recurrence~(\ref{recurAr}) ($H_m$ is not upper-Hessenberg, but this does not affect the derivation).  Also, since the subspaces contain the right-hand side, $r_{0,i}$ is again $V_{m+1}(\beta_i c$) for some $c$, but here $c$ is not a multiple of $e_1$.  Next is the algorithm for the shifting part of the new method that we call GMRES-DR with shifts or GMRES-DR-Sh.  It is the same as for GMRES-Sh, except it uses GMRES-DR instead of GMRES and has the different form of $\bar H_m$.  See~\cite{GMRES-DR} for more on the GMRES-DR portion.  See~\cite{GuZhLi} for a multi-shifted version of the related method GMRES-E~\cite{GMRES-E}.

\vspace{.10in}
\begin{center}
\textbf{GMRES-DR-Sh}
\end{center}
\begin{enumerate}
 \item At the beginning of a cycle of GMRES-DR-Sh, assume the current problem is 
$(A-\sigma_i I) (x_i- \tilde x_{0,i}) = \beta_i r_{0,i},$ with $\beta_1 = 1$, and where $\tilde x_{0,i}$ is the current approximate solution to the i$th$ shifted system. 
 \item Apply GMRES-DR to $A$ and generate Equation~(\ref{recur0}): $AV_m = V_{m+1} \bar H_m$.
 \item For the base system, solve the minimum residual reduced problem $min||c-(\bar H_m - \sigma_1 \bar I)||$, where $c = V_{m+1}^T r_{0,1}$ and $\bar I$ is the $m+1$ by $m$ identity matrix.  The new approximate solution is $\tilde x_1 = \tilde x_{0,1} + V_m d_1$.  The new residual vector is $r_1 = r_{0,1} - A V_m d_1 = r_{0,1} - V_{m+1} \bar H_m d_1$.
 \item For the other shifted systems $i=2, \ldots ns$, form $s = c - (\bar H_m - \sigma_1 \bar I)d_1$.  Apply a QR factorization: $\bar H_m - \sigma_i \bar I = QR$.  Solve $R d_i = \beta_i Q^T c - \beta_i^{new} Q^T s$, using the last row to solve for $\beta_i^{new}$ and the first $m$ rows for $d_i$.
 \item The new approximate solution of the i$th$ system is $\tilde x_i = x_{0,i} + V_m d_i$, and the new residual is $r_i = \beta_i^{new} r_1$.
 \item Test the residual norm for convergence.  If not satisfied, for $i=2 \ldots ns$, set $\beta_i = \beta_i^{new}$ and for $i=1 \ldots ns$, set $\tilde x_{0,i} = \tilde x_i$ and $r_{0,i} = r_i$.  Then go back to step 1.
\end{enumerate} 
\vspace{.15in}

{\it Example 1.} We test the shifted version of deflated GMRES and compare it to shifted regular GMRES.  The matrix has $n=1000$ and is bidiagonal with $0.1, 1, 2, 3, \ldots,$ $998, 999$ on the main diagonal and $1$'s on the superdiagonal.  The right-hand side is generated randomly.  The shifts are $\sigma = 0, -0.4, -2$.  GMRES(25)-Sh is compared with GMRES-DR(25,10)-Sh.  The results are given in Figure 3.1.  This problem has small eigenvalues which slow down restarted GMRES, particularly for the base system with $\sigma = 0$.  Shifting the matrix even by just 0.4 improves convergence of GMRES(25), mainly because the smallest eigenvalue is moved from 0.1 to 0.5.  GMRES-DR converges very rapidly once it generates approximations to the eigenvectors corresponding to these eigenvalues.  The convergence of GMRES-DR for all three shifted systems is similar, since once the small eigenvalue are essentially removed by the deflation, shifting does not have such an important effect.

\begin{figure}
\vspace{.10in}
\begin{center}
\includegraphics[scale=0.5]{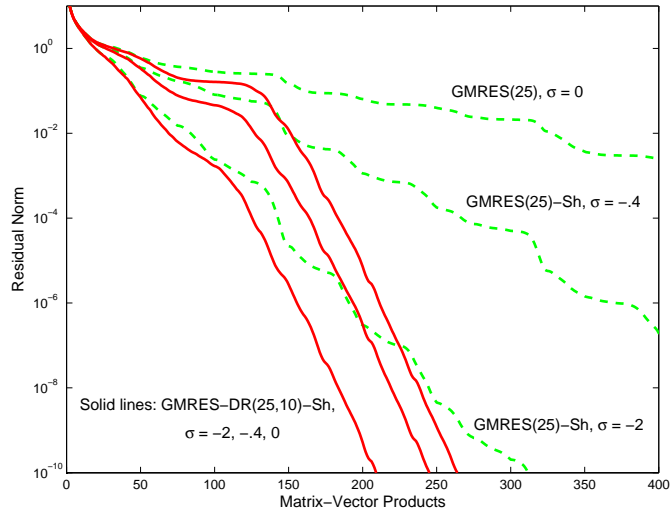}
\end{center}
\vspace{.10in}
\caption{Solution of a system with multiple shifts.}
\end{figure}

In Example 1, the second and third shifted systems converge faster than the first system.  However, in some situations, there can be convergence problems for the non-base systems~\cite{Si03}.  Simoncini~\cite{Si03} compares multiply shifted GMRES and FOM.  Since FOM automatically has residuals parallel for all shifted systems, it can be argued that it is a more natural approach~\cite{Si03}.  However, it is also a matter of where the roots of the FOM and GMRES polynomials fall in relation to the shift.  The next example demonstrates this.  As mentioned earlier, GMRES-Shifts uses the same polynomial for all shifted systems~\cite{FrGl}.  Likewise, FOM-Shifts sticks with one polynomial for all shifts.  The regular GMRES polynomial for an unshifted matrix is scaled to be 1 at zero and needs to be small over the spectrum.  For shifted systems, we have a choice of viewing the polynomial as being 1 at zero and the spectrum shifted or the polynomial being 1 at the shift and the spectrum fixed as that of $A$.  We chose the later.  So for GMRES-Shifts with the base system $A-\sigma_1 I$, we view the polynomial chosen by GMRES as being 1 at $\sigma_1$ and needing to be small over the spectrum of $A$.
In the next two examples, plots are given of the roots of these polynomials.  Because the polynomial needs to be at somewhat small over the spectrum of $A$, a small value of the polynomial at the shift can cause a problem.  With the normalizing, there may then be large values at eigenvalues of $A$.  So for a Krylov method to be effective, the roots of the polynomials need to generally stay away from the shift.

{\it Example 2.}  For the same matrix as in Example 1, we apply GMRES(40)-Sh and FOM(40)-Sh with shifts $\sigma = 0.4, 0$.  The results are shown in Figure 3.2.  For the base shift of $\sigma = 0.4$, GMRES works better, but FOM is more effective on the next shift.  Figure 3.3 shows the harmonic Ritz values~\cite{IE,PaPavdV} nearest the shift for 50 cycles of GMRES.  These are the roots of the GMRES polynomial for the system $(A-0.4I)x_{1}=b$, shifted so they correspond with the spectrum of $A$.  The harmonic Ritz values avoid the region around 0.4, which is good, since the shifted GMRES polynomial for $\sigma = 0.4$ needs to have value 1 at 0.4 and be somewhat small over the spectrum of $A$.  This polynomial may not be effective if it has a root near to 0.4.  GMRES(40) for $\sigma = 0.4$ is able to slowly converge for this fairly difficult, indefinite problem.  Meanwhile, FOM(40) with $\sigma = 0.4$ does not converge, because as shown in Figure 3.3, the roots of the shifted FOM polynomials, which are the Ritz values, are not separated from 0.4.
For the second shift value of 0, GMRES-Sh is not effective.  Too many harmonic Ritz values occur not only near 0, but also on both sides of it.  FOM gives erratic convergence because of some Ritz values near 0, but it does converge.  Although not included in Figure 3.3, we also tested a third shift $\sigma = -1$.  GMRES-Sh again does not converge because of harmonic Ritz values near -1.  This is in spite of the fact that GMRES would have no trouble if this was the base shift.  The Ritz values for FOM do not occur near 1, so FOM converges to under $10^{-8}$ in 400 iterations.  See~\cite{Si03} for more examples in which FOM is better for shifted systems than GMRES.  We next give an example with GMRES more effective.

\begin{figure}
\vspace{.10in}
\begin{center}
\includegraphics[scale=0.5]{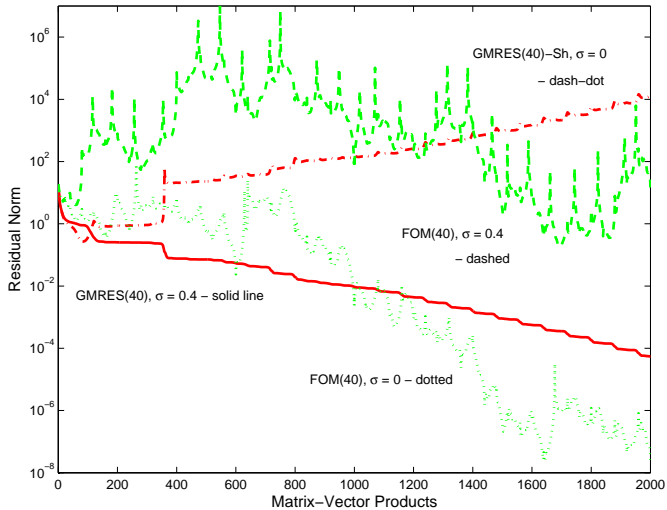}
\end{center}
\vspace{.10in}
\caption{Shifted GMRES and FOM}
\end{figure}

\begin{figure}
\vspace{.10in}
\begin{center}
\includegraphics[scale=0.5]{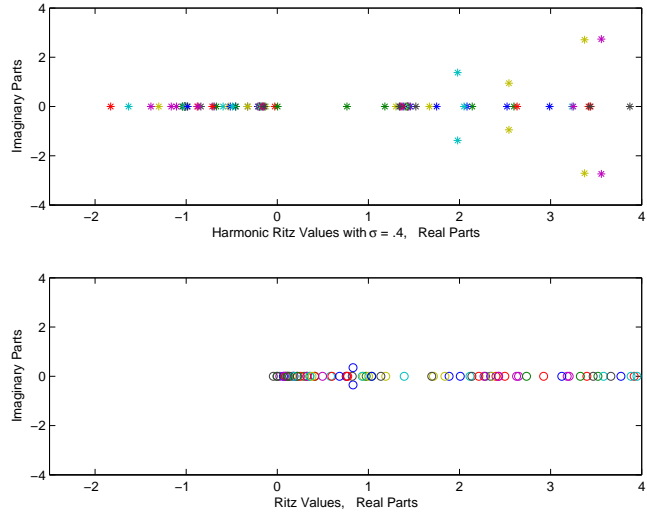}
\end{center}
\vspace{.10in}
\caption{Distribution of Smallest Harmonic and Regular Ritz values, m=40, 50 cycles}
\end{figure}

{\it Example 3.}  The matrix is the same as in the previous example. The base shift is $\sigma_1 = 0$ and a second shift of $\sigma_2 = 1.4$ is used.  GMRES-Sh(80) is compared with FOM-Sh(80).  The results are in Figure 3.4.  Some of the Ritz values fall around 1.4, while there is a gap in the harmonic Ritz values; see Figure 3.5.  So shifted GMRES works much better for the second system.  We conclude that methods for multi-shifted GMRES and FOM must both be used with some caution.  However, deflating eigenvalues can help.  Figure 3.4 also has a plot of FOM-DR-Sh(80,2) for $\sigma = 1.4$, and we see that removing just two eigenvalues fixes the trouble.  For the examples in the rest of this paper, the matrices and shifts are such that the base systems are the ones with the slowest convergence.  

\begin{figure}
\vspace{.10in}
\begin{center}
\includegraphics[scale=0.5]{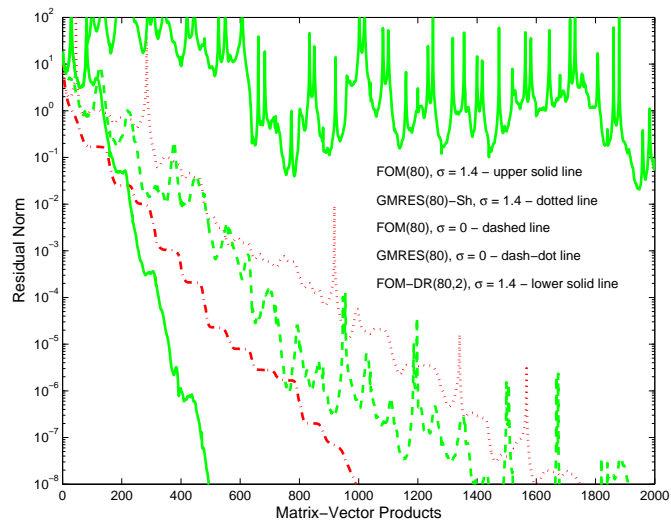}
\end{center}
\vspace{.10in}
\caption{GMRES-Sh versus FOM-Sh}
\end{figure}

\begin{figure}
\vspace{.10in}
\begin{center}
\includegraphics[scale=0.5]{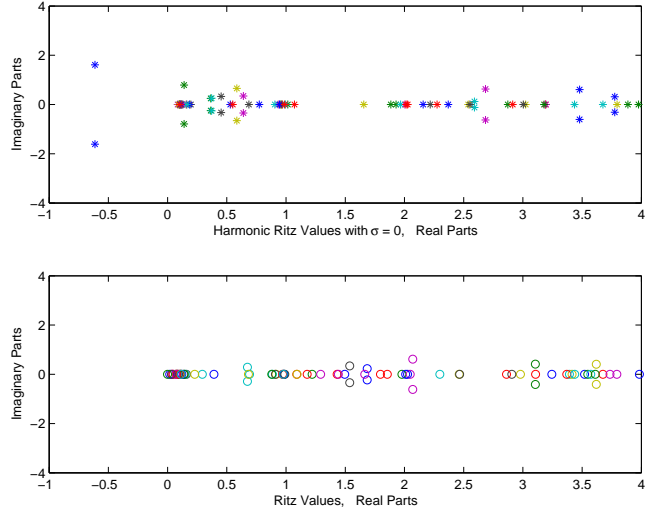}
\end{center}
\vspace{.10in}
\caption{Distribution of Smallest Harmonic and Regular Ritz values, m=80, 25 cycles}
\end{figure}

\section{Deflated GMRES for multiple right-hand sides and multiple shifts}

We now consider solving multiply shifted systems that also have multiple right-hand sides.  It is important to reuse information or share information among the right-hand sides.  It is possible to design multi-shifted versions of both Block-GMRES~\cite{Sa96} and Block-GMRES-DR~\cite{bgdr}.  However, here we will concentrate on a non-block approach.  The right-hand sides are solved separately, and eigenvector information from solution of the first right-hand side is used to assist the subsequent ones.  More specifically, we will generalize for multiple shifts the GMRES-Proj approach mentioned in Section 2.  See~\cite{gproj} for more on this method, including comparison with block methods.

First some of the difficulties of deflating for subsequent right-hand sides will be discussed.  Suppose the first right-hand side has been solved and approximate eigenvectors have been generated.  Then for the non-shifted case, there are several ways to deflate eigenvalues.  Some of these are given in~\cite{gproj}.  However, generally they do not work for multiply shifted systems.  For example, some deflation approaches involve building a preconditioner from the approximate eigenvectors~\cite{BaCaGoRe,BuEr,gproj}.  As mentioned earlier, shifted systems cannot be solved together if there is preconditioning.  

For the GMRES-Proj method, there is trouble with one of the two phases.  We know the GMRES portion can be adapted to keep right-hand sides parallel for multiple shifts.  However, the phase with projection over approximate eigenvectors generally fails to produce parallel residual vectors.  Even though this projection is over a Krylov subspace of dimension $k$ spanned by the columns of $V_k$, this subspace does not contain the current right-hand side (the residual vector).  So the derivation in Section 2.1 does not work with $V_{k+1}$ and $\bar H_k$ from (\ref{recur1}) replacing $V_{m+1}$ and $\bar H_m$ from (\ref{recurAr}).  Specifically, the transition from Equation~(\ref{gsh1}) to Equation~(\ref{gsh2}) is not possible since $r_{0,i}$ and $r_{0,1}$ are not in the span of the columns of $V_{k+1}$.  One case where the projection does keep the residual vectors parallel is with exact eigenvectors.  We now show this.

\begin{theorem}
Assume that before the minres projection, the shifted systems are \[(A-\sigma_i I)(x_i-\tilde x_{0,i}) = r_{0,i}\,\] 
with $r_{0,i} = \beta_i r_{0,1}$ for $i = 2, \ldots, ns$.  Let $z_1, z_2, \ldots, z_k$ be eigenvectors of $A$.
Then after the minres projection over the subspace $Span\{z_1, z_2, \ldots, z_k\}$, the residual vectors are parallel.
\end{theorem}

{\em Proof}.  Let $Z_k$ be the matrix with $z_1, \ldots, z_k$ as columns.  For exact eigenvectors, the minres projection is equivalent to Galerkin~\cite{Sa96}.  The Galerkin projection gives the reduced problem: 
\[Z_k^T(A-\sigma_i I) Z_k d_i = \beta_i Z_k^T r_{0,1}. \]
Solving gives 
\[d_i = \beta_i (\Lambda_k - \sigma_i I_k)^{-1} (Z_k^TZ_k)^{-1}Z_k^T r_{0,1}, \]
where $\Lambda_k$ is the $k$ by $k$ diagonal matrix with diagonal entries $\lambda_1$ through $\lambda_k$ and $I_k$ is the $k$ by $k$ identity matrix.
The residual vector after projecting is then 
\begin{eqnarray*}
r_{i} &=& \beta_i r_{0,1} - (A-\sigma_i I) Z_k d_i \\
&=& \beta_i r_{0,1} - \beta_i (A-\sigma_i I)Z_k (\Lambda_k - \sigma_i I_k)^{-1} (Z_k^TZ_k)^{-1}Z_k^T r_{0,1} \\
&=& \beta_i r_{0,1} - \beta_i Z_k(\Lambda_k - \sigma_i I_k) (\Lambda_k - \sigma_i I_k)^{-1} (Z_k^TZ_k)^{-1}Z_k^T r_{0,1}  \\
&=& \beta_i (I - Z_k (Z_k^TZ_k)^{-1}Z_k^T) r_{0,1}.  
\end{eqnarray*}
This shows that all $r_{i}$ are multiples of each other.
\endproof

So one option for GMRES-Proj with multiple shifts is to use only fairly accurate eigenvectors.  We could sort through the approximate eigenvectors computed by GMRES-DR and apply only ones with acceptible accuracy to the projection in GMRES-Proj. This is now tested.  

{\it Example 4.} For the same matrix as in Example 1, we first solve the $\sigma_1 = 0$ system to accuracy of relative residual norm below $10^{-10}$.  This takes 250 matrix-vector products.  The second column of Table 4.1 shows the accuracy of the $k$th eigenvectors thus produced.  For example, the residual norm of the tenth approximate eigenvectors is 4.1e-2, while the fourth is much better at 1.0e-6.  Now we consider solution of the second right-hand side using the first $k$ eigenvectors with $k$ from 10 down to 2.  The third column gives the number of matrix-vector products for the relative residual norm of the base shifted system with the second right-hand side to reach 1.e-10.  We see that convergence is better using all ten eigenvectors.  However, the fourth column gives the acccuracy attained by the worst of the last two shifted systems (usually it is the third shift).  It only reaches residual norm of 3.3e-4 if all 10 approximate eigenvectors are used in the projection.  With only four eigenvectors, the residual norm reaches a better level of 5.4e-8, but the convergence is almost twice as slow.  

The last three columns repeat this information for the case of solving the first right-hand side system to greater accuracy of relative residual norm of 1.e-14 (385 matrix-vector products).  The second right-hand side systems are then able to be solved more accuracy.  Even with $k=8$ eigenvectors, all shifted systems reach accuracy of 1.3e-9 or better compared to 3.7e-5 for the previous case of less accurate eigenvectors.

\begin{table}
\caption{Effect of projecting over different accuracies of eigenvectors} 

\begin{center} \footnotesize
\begin{tabular}{|c|c|c|c|c|c|c|c|}  \hline\hline
       & 250 mvp's for 1st   &   &         & 385 mvp's  1st & &    \\ \hline
$k$ & eig. res. & mvp's  & lin. eqs. res.  &  eig. res.  & mvp's  & lin. eqs. res.  \\  
\hline \hline
10  & 4.1e-2  & 135    & 3.3e-4  & 4.4e-1  & 135 & 1.7e-6 \\ \hline
8   & 3.3e-3  & 165    & 3.7e-5  & 7.1e-6  & 165 & 1.3e-9 \\ \hline
6   & 8.0e-5  & 180    & 2.8e-6  & 3.3e-10 & 180 & 1.3e-9 \\ \hline
4   & 1.0e-6  & 255    & 5.4e-8  & 2.4e-11 & 255 & 5.5e-10 \\ \hline 
2   & 5.4e-9  & 435    & 1.5e-8  & 9.9e-12 & 435 & 3.6e-10 \\ \hline

\hline\hline 

\end{tabular} 
\end{center} 
\end{table}

The problem with this approach is that the eigenvector computation during solution of the first right-hand side needs to be done to considerable accuracy, since we do not want to slow down convergence of the subsequent systems.  If many right-hand sides are to be solved, this extra expense might not be significant.  However, we next propose an approach without this concern of needing accurate eigenvectors.  

The key idea is that although the residual vectors cannot be kept parallel, they can be chosen so that they relate to each other.  We force the residuals of the non-base systems to be parallel to the residual of the base system except for a component in the direction of $v_{k+1}$, the last column of the $V_{k+1}$ matrix from Equation~(\ref{recur1}).  We then continue solving but ignore this component.  At the end, a correction can be done.  However, for this correction, we need solution of shifted systems with one additional right-hand side, namely $v_{k+1}$.  We begin discussion of this approach with the aspect of keeping residuals parallel except for the one component, then move on to the correction phase.

Suppose we have shifted systems with parallel residuals, namely $(A-\sigma_i I)
(x_i- \tilde x_{0,i}) = \beta_i r_{0,1},$ with $\beta_1 = 1$.  A projection over the columns of $V_k$ can be implemented to give for the base system a new approximate solution $x_1$ and residual $r_1$ and for the other systems new approximate solutions $x_i$ and residuals $\beta_i r_1$ such that $(A-\sigma_1 I)(x_1- \tilde x_1) = r_{1},$ and 
\begin{equation}
(A-\sigma_i I)(x_i- \tilde x_i) = \beta_i r_{1} + \gamma_i v_{k+1}, \label{keyder1}
\end{equation}
for $i=2,\ldots, k$.
First apply minres projection over the approximate eigenvectors spanned by the columns of $V_k$ to the base system.  The base system residual vector is then $r_1 = r_{0,1} - AV_{k+1} d_1$.  We need $d_i$ to be chosen so that $r_i = \beta_i r_1 + \gamma_i v_{k+1}$, for some scalar $\gamma_i$.  So we need
\[ 
r_{0,i} - (A-\sigma_i I) V_k d_i = \beta_i (r_{0,1} - (A-\sigma_1 I)V_k d_1) + \gamma_i v_{k+1}. \nonumber 
\] 
After using that $r_{0,i} = \beta_i r_{0,1}$ and using the key recurrence for the approximate eigenvectors~(\ref{recur1}), we have
\[
V_{k+1} (\bar H_k -\sigma_i \bar I) d_i = \beta_i V_{k+1} (\bar H_k -\sigma_1 \bar I)d_1 - \gamma_i v_{k+1}.
\]
Next the $V_{k+1}$ can be dropped, and we let $t = \beta_i (\bar H_k - \sigma_1 \bar I)d_1$:
\begin{equation}
(\bar H_k - \sigma_i \bar I) d_i = t - \gamma_i e_{k+1}, \label{keyder2}
\end{equation}
where $e_{k+1}$ is the $k+1$st coordinate vector of length $k+1$.
We ignore the last row of (\ref{keyder2}) and solve the first $k$ equations for the unknown vector $d_i$ of length $k$.  Then Equation~(\ref{keyder2}) is automatically true for some $\gamma_i$.  So we have (\ref{keyder1}).  Note the $\beta_i$'s do not change during this projection over approximate eigenvectors, unlike in the GMRES-Sh portion.

Now we will look at the correction phase that is needed at the end of GMRES-Proj.  Assume that we have already solved shifted systems with the extra right-hand side $v_{k+1}$ (this solution will be discussed next) and have 
\begin{equation}
(A-\sigma_i I)s_i = v_{k+1}. \label{keyder3}
\end{equation}
We assume that for a particular right hand side, the systems have been solved by GMRES-Proj to the point that the residual is only in the direction of $v_{k+1}$ and the system is recast as
\begin{equation}
(A - \sigma_i I) (x_i - \tilde x_i) = r_i = \gamma_i v_{k+1}, \label{keyder4}
\end{equation}
for $i=2,\ldots,ns$ and for some scalar $\gamma_i$.  Here $\tilde x_i$ is the approximate solution to $x_i$.  We perform a Galerkin projection for system~(\ref{keyder4}) over the subspace spanned by the single vector $s_i$ from solution of~(\ref{keyder3}): 
\[s_i^T(A-\sigma_i I)s_i \delta = \gamma_i s_i^T v_{k+1}.\]  Using Equation~(\ref{keyder3}), this becomes \[s_i^T v_{k+1} \delta = \gamma_i s_i^T v_{k+1}.\]  Then $\delta = \gamma_i$.  To determine $\gamma_i$, we start with $r_i = \gamma_i v_{k+1}$. Multiplying both sides by $v_{k+1}^T$ and using that $v_{k+1}$ is of unit length gives \begin{equation}
\gamma_i = v_{k+1}^T r_i. \label{keyder4b}
\end{equation}
The corrected solution of the system is $x_i = \tilde x_i + \delta s_i = \tilde x_i +  v_{k+1}^T r_i s_i$.

We need to fill in the method for solving~(\ref{keyder3}), the shifted systems with the extra right-hand side $v_{k+1}$.  First, GMRES-Proj is applied.  This includes projection over the approximate eigenvectors as described above alternating with cycles of GMRES.   This continues until the residual is negligible except in the direction of $v_{k+1}$.  So for the correction, we assume that
\begin{equation}
(A - \sigma_i I) \tilde s_i = r \label{keyder5}
\end{equation}
is the current system, where
\[
r = v_{k+1} - (A - \sigma_i I) \tilde s_i = \gamma_i v_{k+1},
\]
for some scalar $\gamma_i$.
Rearranging this residual equation gives 
\begin{equation}
(A - \sigma_i I) \tilde s_i = (1-\gamma_i) v_{k+1}.\label{keyder6}
\end{equation}
Applying Galerkin projection over the subspace spanned by the single vector $\tilde s_i$ to the system (\ref{keyder5}) gives
\[
\tilde s_i^T (A - \sigma_i I) \tilde s_i \delta = \gamma_i \tilde s_i^T v_{k+1}.
\]
With Equation~(\ref{keyder6}), this becomes
\[
(1-\gamma_i) \tilde s_i^T v_{k+1} \delta = \gamma_i \tilde s_i^T v_{k+1},
\]
and this simplifies to
\[\delta = {\gamma_i \over {1-\gamma_i}}.\] 
So the corrected solution is 
$s_i = \tilde s_i + {\gamma_i \over {1-\gamma_i}} \tilde s_i = { 1\over {1-\gamma_i}} \tilde s_i.$
Finally, the $\gamma_i$ can be determined to be $\gamma_i = v_{k+1}^T r_i $ as it was for (\ref{keyder4b}).

We next list the algorithms for solution of the systems with second and subsequent right-hand sides and with the extra right-hand side.  They are given in the order they were derived here, not in order of how they are actually used.

\vspace{.10in}
\begin{center}
\textbf{GMRES-Proj-Sh for the second and subsequent right-hand sides}
\end{center}
\begin{enumerate}
 \item Consider the systems with the $j$th right-hand side (and with all $ns$ shifts). 
 \item At the beginning of a cycle of GMRES-Proj-Sh, assume the current problem is 
$(A-\sigma_i I) (x_i- \tilde x_{0,i}) = \beta_i r_{0,i},$ with $\beta_1 = 1$, and where $\tilde x_{0,i}$ is the current approximate solution to the i$th$ shifted system.
 \item Apply the Minres Projection for $V_k$ to the base shift.  This uses the $V_{k+1}$ and $\bar H_k$ matrices developed while solving the first right-hand side with GMRES-DR-Sh.  
 \item  For shifted systems $is=2 \ldots ns$, solve $(H_k - \sigma_i I) d_i = \beta_i (H_k - \sigma_1 I)d_1$, where $H_k$ is the $k$ by $k$ portion of $\bar H_k$.  Set the new approximate solution as $\tilde x_{i} = \tilde x_{0,i} + V_k d_i$.
 \item Apply one cycle of GMRES(m)-Sh.
 \item Test the residual norms for convergence (can also test during the GMRES cycles).  For the non-base systems, ignore the error term in the direction of $v_{k+1}$.  Residual formula~(\ref{gsh3}) can be used.  If not satisfied, go back to step 2.  Otherwise conclude with step 7.
 \item Correction phase: Suppose the computed solution so far for the $i$th shifted system is $\tilde x_i$.  Let the solution to the system with the extra right hand side $v_{k+1}$ and shift $\sigma_i$ be $s_i$.  The corrected solution is $x_i = \tilde x_i + (v_{k+1}^T r) s_{i}$ and the corrected residual norm can be calculated.
\end{enumerate} 
\vspace{.15in}

\vspace{.10in}
\begin{center}
\textbf{GMRES-Proj-Sh for the extra right-hand side $v_{k+1}$}
\end{center}
Same as for previous algorithm except for
\begin{enumerate}
 \item Consider the systems with right-hand side $v_{k+1}$ (and with all $ns$ shifts). 
 \setcounter{enumi}{6} 
 \item Correction phase: Suppose the computed solution for the $i$th shifted system so far is $\tilde s_i$.  The corrected solution is $s_i = ({ 1\over {1-\gamma_i}})*\tilde s_{i}$, with $\gamma_i = v_{k+1}^T r.$
\end{enumerate} 
\vspace{.15in}

{\it Example 5.}  We use the same test matrix.  All right-hand sides are generated randomly.  The systems with the first right-hand side are solved with GMRES-DR(25,10)-Sh as in Example 1.  Then the extra right-hand $v_{k+1}$ systems are solved (for all shifts) with GMRES(15)-Pr(10)-Sh.  Finally, the second right-hand side systems are also solved with GMRES(15)-Pr(10)-Sh.  All relative residual tolerances are $rtol=1.e-6$.  Figure 4.1 has residual curves for only two shifts, the base shifts of zero and $\sigma = -2$.  The solid line is the base shift.  The dotted line shows the uncorrected residual norm for the second shift, while the dash-dot line has the second shift residuals if they are corrected (actually the correction needs to be done only once at the end).  The uncorrected residual norm for the second right-hand side and second shift levels off at 4.e-3, but this is fixed by the correction phase.  The convergence is faster than for GMRES-DR-Sh, because the eigenvectors are used from the beginning to speed up the convergence.  Also the cost of GMRES(15)-Proj(10)-Sh is less than for GMRES-DR(25,10)-Sh, because it is fairly inexpensive to project over the approximate eigenvectors compared to keeping the eigenvectors in the GMRES-DR subspace.  Here the expense for the extra right-hand side is fairly significant, however it will not be if there are more right-hand sides.  Figure 4.2 has the case of solving a total of 10 right-hand sides.  Also, the extra right-hand side is solved only to relative residual tolerance of 1.e-3.  Now the expense for the extra right-hand side $v_{k+1}$ is small compared to amount saved by speeding up solution of all the right-hand sides.

\begin{figure}
\vspace{.10in}
\begin{center}
\includegraphics[scale=0.5]{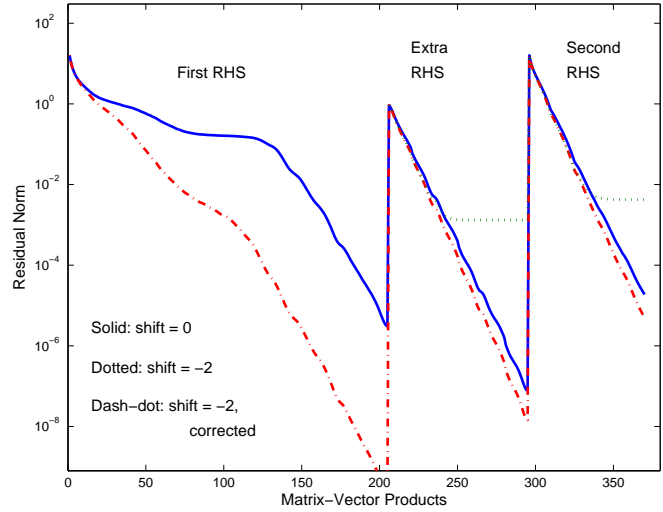}
\end{center}
\vspace{.10in}
\caption{Solution of first rhs, extra rhs and second rhs with two shifts.}
\end{figure}

\begin{figure}
\vspace{.10in}
\begin{center}
\includegraphics[scale=0.5]{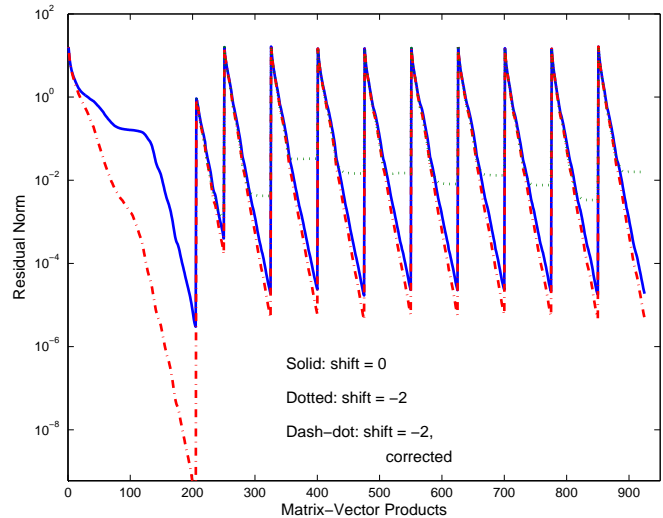}
\end{center}
\vspace{.10in}
\caption{Solution of ten right-hand sides with two shifts.}
\end{figure}

{\it Example 6.}
At the end of the previous example, the extra right-hand side is solved to low accuracy, but the correction for the subsequent right-hand sides is still successful.
We now experiment with solving the extra right-hand side to different levels of accuracy.  Table 4.2 shows the accuracy after correction for the $\sigma=-2$ system when the extra right-hand side system is solved to relative residual tolerances ranging from 1.e-6 down to 1.e-1 (the tolerance is checked for termination only at the end of GMRES cycles).  The first and second right-hand side systems are solved to three different relative residual norm tolerances (1.e-6, 1.e-8 and 1.e-10) in the three rows of the table.  The conclusion of this experiment is that the extra right-hand side systems do not need to be solved very accurately.  For instance, with desired  tolerance 1.e-8 for the first and second right-hand sides, the extra right-hand side systems need only to be solved to relative tolerance of 1.e-3 to get accuracy of 2.5e-7 (and relative accuracy of 7.9e-9) for the second right-hand side system.

\begin{table}
\caption{Effect of solving the extra right-hand side system to different accuracies} 

\begin{center} \footnotesize
\begin{tabular}{|c|c|c|c|c|c|c|c|c|}  \hline\hline
desired rtol  	& accurracy of $2$nd    &   & & & & &  \\ 
of $1$st and $2$nd sys's	 & before correction & 1.e-6  & 1.e-5  &  1.e-4  & 1.e-3  & 1.e-2 & 1.e-1  \\   
\hline \hline
1.e-6 & 4.2e-3  & 4.8e-6  & 4.8e-6    & 4.8e-6  & 4.9e-6  & 6.4e-6 & 3.8e-4 \\ \hline
1.e-8 & 3.6e-4  & 2.4e-8  & 2.4e-8    & 2.4e-8  & 2.5e-7  & 9.4e-7 & 9.9e-5 \\ \hline
1.e-10 & 1.2e-3 & 1.5e-10 & 2.7e-10   & 1.0e-9  & 9.7e-8  & 2.7e-7 & 3.1e-5 \\ \hline

\hline\hline 

\end{tabular} 
\end{center} 
\end{table}  

The next example is probably the key example in the paper.  It shows the value of deflating eigenvalues for an important application.  QCD problems often have the need of solution of multiple right-hand sides and multiple shifts for each right-hand side.  They also have complex spectra with small eigenvalues for the problems of most interest.

{\it Example 7.}
We look at a Wilson-Dirac matrix from lattice QCD~\cite{Qcdconf}.  The matrix is complex and the dimension is 393,216 by 393,216.  The value of $\kappa$ is 0.158 for the base shift.  This is approximately $\kappa_{critical}$.  The eigenvalues are in the right-half of the complex plane, but partially surround the origin~\cite{Qcdconf}.  The right-hand sides are unit vectors associated with particular space-time, Dirac and color coordinates.  Often there are a dozen or more right-hand sides associated with each matrix and perhaps seven shifts for each right-hand side.  We will just show solution of the second right-hand side for three shifts, $\sigma = 0, -0.3, -0.5$.
The first right-hand side is solved with GMRES-DR(50,30) to residual tolerance of 1.e-8 and the extra right-hand side to 1.e-7.  Then for the second right-hand side, GMRES-Proj uses 30 approximate eigenvectors for the projection in between cycles of GMRES(20).  See Figure 4.3 for the results.  GMRES(20)-Proj(30)-Sh converges in about one-tenth of the iterations needed for GMRES(20).  To reach residual norm of less than $10^{-7}$ for the toughest system with shift of zero takes 2680 matrix-vector products for GMRES(20)-Sh and 280 for GMRES(20)-Proj(30)-Sh.

\begin{figure}
\vspace{.10in}
\begin{center}
\includegraphics[scale=0.5]{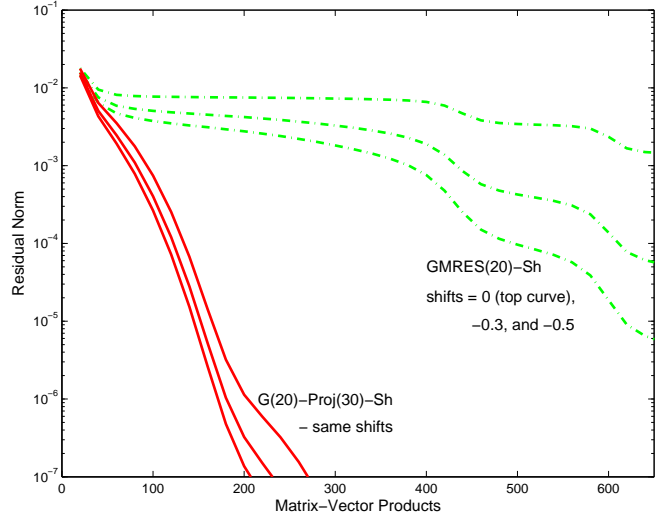}
\end{center}
\vspace{.10in}
\caption{Solution of ten related right-hand sides.}
\end{figure}

\section{Related Right-hand Sides}

We need to take advantage of any relationship between the right-hand sides.  This section shows that this can easily be done even for the case of multiple shifts. 
 
Assume we have solutions of all shifted systems for right-hand sides 1 through $j$.  So we have 
\[
(A - \sigma_i I)x_i^{irhs} = b^{irhs},
\] 
for $irhs = 1, \ldots, j-1$.
We put these equations together to form 
\begin{equation}
(A - \sigma_i I)X_i = B, \label{relrhs1}
\end{equation} 
where $B$ is the $n$ by $j-1$ matrix with columns $b^1$ through $b^{j-1}$, and $X_i$ has columns $x_i^1$ through $x_i^{j-1}$.
We assume there is no initial guess.  Applying Minres projection over the subspace spanned by the columns of $X_i$ to the system with the right-hand side $j$ and shift $\sigma_i$ gives 
\[
X_i^T (A - \sigma_i I)^T (A - \sigma_i I)X_i d = X_i^T (A - \sigma_i I)^T b^{j}.
\]
With (\ref{relrhs1}), this becomes 
\begin{equation}
B^T B d = B^T b^{j}.  \label{relrhs2}
\end{equation}
Note that the solution of Equation (\ref{relrhs2}) is independent of the shift.  This makes the residual vectors all the same: 
\[r_i = b^{j} - Bd.\]  The approximate solutions are \[\tilde x_i^j = X_i d.\]  So this approach projects over all of the previous solutions for each shifted system and provides the needed parallel residuals. 

{\it Example 8.}  We repeat the test in Example 5 with 10 right-hand sides, except this time they are related to each other.  We define the second and subsequent right-hand sides as $b^j = b^1 + 10^{-4}*u^j$, where $u^j$ is a random vector (both $b^1$ and $u^j$ have elements distributed normally with mean 0 and variance 1).  
Before solving the second and subsequent right-hand sides, we project over the previous solutions as just described.  The results are in Figure 5.1.  Using the close relationship between the right-hand sides allows the number of matrix-vector products to be cut in half for each of the subsequent right-hand sides.

\begin{figure}
\vspace{.10in}
\begin{center}
\includegraphics[scale=0.5]{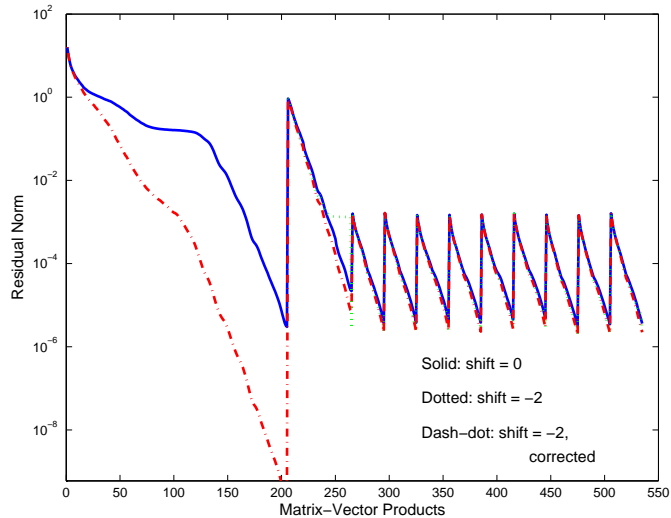}
\end{center}
\vspace{.10in}
\caption{Solution of first rhs, extra rhs and second rhs with two shifts.}
\end{figure}

\section{Conclusion}

Deflating eigenvalues can significantly improve restarted GMRES for matrices with small eigenvalues.  This work focuses on deflated GMRES for the case of multiple shifts of the matrix.
When using Krylov methods to solve systems of equations with multiple shifts, the goal is to solve all systems with about the same expense as one system.  Past work has developed such methods for non-restarted and even restarted Krylov methods.   Here this is extended for deflated, restarted GMRES methods, both in the case of a single right-hand side and for multiple right-hand sides.   
For multiple right-hand sides, there is added expense for solving an extra right-hand side, but this allows all shifted systems to be solved simultaneously for as many right-hand sides as are desired.  Also, it is possible to efficiently take advantage of closely related right-hand sides, even in this case of multiple shifts.
These approaches can be very beneficial for an important application in lattice QCD physics.  

Future work could examine other QCD problems including the overlap fermion problem, which involves solving an inner-outer loop.  There may be potential for deflation in both of the loops.  Deflation of non-restarted methods such as BiCGStab should also be investigated.  Another interesting topic would be solution of systems with changing shifts, such as may occur in model order reduction~\cite{GuAnBe}.

\bibliography{morgan}

\end{document}